\documentclass{iaus}
\usepackage{graphicx}

\newcommand{\Ha}{H$\alpha$}
\newcommand{\Hb}{\ifmmode {\rm H}\beta \else H$\beta$\fi}

\newcommand{\Nii}{[N\,{\sc ii}]\,$\lambda$6584}

\newcommand{\Oii}{[O\,{\sc ii}]\,$\lambda$3727}

\newcommand{\Oiii}{[O\,{\sc iii}]\,$\lambda$5007}
\newcommand{\Oiiit}{[O\,{\sc iii}]\,$\lambda$4363}

\newcommand{\Neiii}{[Ne\,{\sc iii}]\,$\lambda$3869}

\newcommand{\Sii}{[S\,{\sc ii}]\,$\lambda$6716,\,$\lambda$6731}

\newcommand{\Siii}{[S\,{\sc iii}]\,$\lambda$9069}

\newcommand{\Ariii}{[Ar\,{\sc iii}]\,$\lambda$7135}

\title[Nebular abundances in galaxies] 
{Nebular abundances in galaxies:\\            Beware of biases}

\author[Gra\.zyna Stasi\'nska]   
{Gra\.zyna Stasi\'nska$^1$}

\affiliation{$^1$LUTH, Observatoire de Paris, CNRS, Universit\'e Paris Diderot; Place Jules Janssen 92190 Meudon, France \\email: {\tt grazyna.stasinska@obspm.fr}}

\pubyear{2009}
\volume{xxx}  
\pagerange{ }
\jname{Stellar Populations - Planning for the Next Decade}
\editors{G. Bruzual,  \& S. Charlot, eds.}
\begin{document}

\maketitle

\begin{abstract}
The derivation of nebular abundances in galaxies using strong line methods is simple and quick. Various indices have been designed and calibrated for this purpose, and they are widely used. However, abundances derived with such methods may be significantly biased, if the objects under study have different structural properties (hardness of the ionizing radiation field, morphology of the nebulae) than those used to calibrate the methods.  Special caution is required when comparing the metallicities of different samples, like, for example, blue compact galaxies and other emission line dwarf galaxies, or samples  at different redshifts. 
\end{abstract}

\firstsection 
\section{Introduction}

Why talk about nebular abundances in a symposium devoted to stellar populations in galaxies? The ultimate goal of stellar populations studies is to understand the evolution of galaxies.  In principle, the determination of  the metallicities of the stellar populations gives an evolutionary view of the chemical enrichment of galaxies. Such studies are just beginning (see e.g. Vale Asari et al. 2009 and contribution to this symposium).  For the moment, however, the overwhelming majority of studies on the chemical composition of galaxies  rely on global stellar metallicity indices (for early-type galaxies),  and on nebular abundances (for star-forming galaxies) derived from emission-line studies. Note that the abundances derived by one or the other method do not have the same meaning: Stellar indices provide an integrated (and weighted) metallicity of the different generations of stars while the chemical composition of nebulae is the result of mixing with the interstellar gas of the metals ejected by \emph{all} the present and past stellar generations and is the same as that of the presently forming stars.

In this communication, we would like to draw attention to widely overlooked biases in the determination of nebular abundances. We will restrict to the so-called ``strong line methods'' for abundance determination (see e.g. Stasi\'nska 2004 for a general introduction on nebular abundance determinations). After presenting a number of abundance calibrators,  We will show how their use may, in certain cases,  lead to erroneous statements.
 
\section{A summary of strong line methods and their recent calibrations}

Strong line methods are statistical in nature, and take advantage of the fact that, in giant HII regions,  the metallicity appears to be linked to the mean effective temperature of the stellar radiation field and to the ionization parameter. They are presently being applied to integrated spectra of galaxies (e.g. Kewley \& Ellison 2008), which poses the yet unsolved problem of how the galactic  abundance gradients affect the results. Here, we will ignore this problem, and show that abundances obtained by strong line methods can be biased even in the case of chemically homogeneous metal-poor galaxies and, therefore, lead to incorrect interpretations.

Strong line methods, being statistical, have to be calibrated. This can be done either using a grid of photoionization models or a sample of objects for which the oxygen abundances have been determined independently by \emph{direct} methods. 

Here is a list of abundance calibrators which are routinely used for nebular abundance studies.

\begin{itemize}

\item  \textbf{a}  O$_{23}$, i.e. (\Oii\ + \Oiii)/\Hb, introduced by Pagel et al. (1979)
\item \textbf{b}  O$_{3}$N$_{2}$, i.e. \Oiii/\Nii, proposed by Alloin et al. (1979)
\item \textbf{c}  S$_{23}$, i.e. (\Sii\ + \Siii)/\Hb, proposed by Vilchez \& Esteban (1996)
\item \textbf{d}  N$_{2}$, i.e.\Nii/\Ha, first used by Storchi Bergmann et al. (1994)
\item \textbf{e}  S$_{3}$O$_{3}$, i.e. \Siii/\Oiii, proposed by Stasi\'nska (2006)
\item \textbf{f}  Ar$_{3}$O$_{3}$, i.e. \Ariii/\Oiii, proposed by Stasi\'nska (2006)
\item \textbf{g}  Ne$_{3}$O$_{2}$, i.e. \Neiii/\Oiii, proposed by Nagao et al. (2006)
\item \textbf{h}  O$_{3}$O$_{2}$, i.e. \Oiii/\Oii, considered by Nagao et al. (2006)

\end{itemize}

As emphasized by Stasi\'nska (2008), not all of those indices have a \emph{direct} link with the metallicity (or rather with the oxygen abundance, which is taken as a proxy for metallicity in nebular studies). It is the empirical relation between metallicity and structural properties of HII regions that allows them to be used as metallicity indicators. For example, the hardness of the ionizing radiation in giant HII regions is correlated with metallicity,  due to opacity effects in stellar interiors atmospheres that soften the ionizing radiation field (Mc Gaugh 1991). The ionization parameter is correlated with metallicity  due to  stellar wind pressure that shapes the morphology of the emitting gas (Dopita et al 2006). Finally, N/O is related to O/H due to secondary production of nitrogen.

Those indices (and others) have been calibrated many times. Here, we will use the calibration by Nagao et al. (2006) for  O$_{23}$, O$_{3}$N$_{2}$, N$_{2}$, Ne$_{3}$O$_{2}$ and  O$_{3}$O$_{2}$,  P{\'e}rez-Montero \& D{\'{\i}}az (2005) for S$_{23}$ and Stasi\'nska (2006) for S$_{3}$O$_{3}$ and Ar$_{3}$O$_{3}$.

\section{Evidence for biases with strong line methods}

The following is motivated by the strange result of S{\'a}nchez Almeida et al. (2008) that BCDs (blue compact dwarf galaxies) in the Sloan Digital Sky Survey (SDSS) have, on average, lower oxygen abundances than QBCDs (quiet blue compact dwarf galaxies). This claim is based on abundances derived using the N$_{2}$ index.

\begin{figure*}[t]
\begin{center}
 \includegraphics[width=\textwidth]{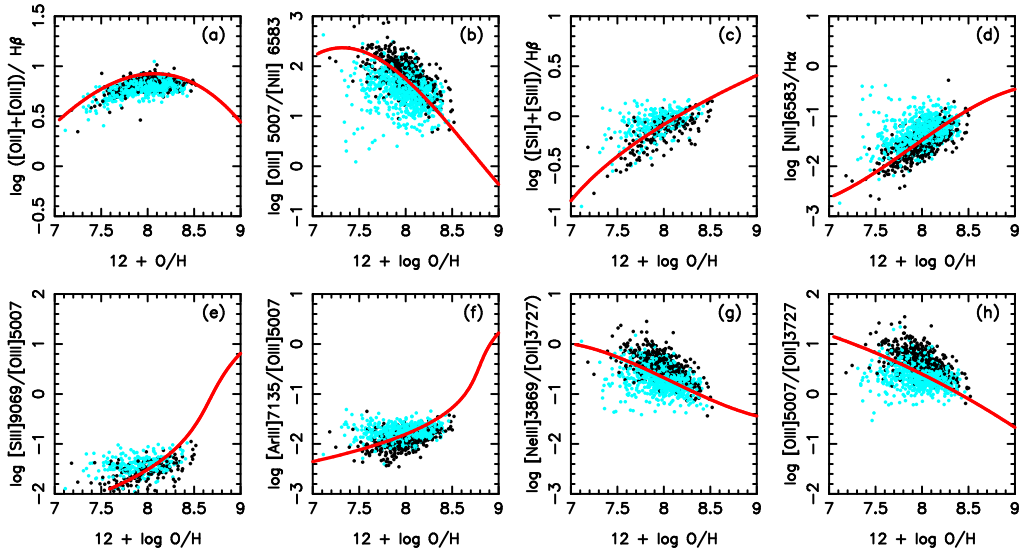} 
 \caption{The values of some strong-line ratios used as abundance indicators, as a function of the oxygen abundance. The curves represent various calibrations (as indicated in the text). The dots represent galaxies from the SDSS for which O/H was derived with a temperature-based method.  Black dots represent galaxies with $W_{\Hb} > 100$\AA. Cyan dots represent galaxies with $W_{\Hb} < 100$\AA. Clearly, strong line methods tend to underestimate the values of O/H for black points and overestimate them for cyan points.}
   \label{fig1}
\end{center}
\end{figure*}

That oxygen abundances derived from methods involving nitrogen lines may be biased has been noted many times. We will show below that the problem is more serious than that and affects any statistical method.   
We consider all the emission line galaxies from the Sloan Digital Sky Survey (data release 6, Adelman-McCarthy et al. 2008) for which the weak \Oiiit\ line has been measured with better than 30\% accuracy, and divide this sample in two groups. One which has H$\beta$ equivalent widths, $W_{\Hb}$, larger than 100\AA, the other with   smaller values of $W_{\Hb}$. The first group bears similarities with the BCD class of S{\'a}nchez Almeida et al. (2008) (newborn galaxies), while the second group contains objects more like their QBCDs (aging galaxies), since low values of  $W_{\Hb}$ tend to indicate more evolved populations of the ionizing stars. For all those objects, we can obtain the oxygen abundance with the direct, temperature-based method, following the recipes given in Izotov et al. (2006). We can also compute the oxygen abundances using strong line indices listed above. In the various panels of Fig. 1, we plot the values of those indices as a function of O/H as derived from the direct method. BCDs are represented by black points, QBCDs by cyan (light grey) points. The calibrating functions presented in the references  at the end of Sect. 2 are shown as continuous lines. 

The result is eloquent. QBCDs are \emph{vertically} displaced from BCDs in all the diagrams. There is no indication of a difference in O/H values (derived with direct methods) between the two groups. On the other hand, if one were to use the strong line indices, 
Fig. 1 clearly shows that the O/H values would tend to be underestimated for BCDs and overestimated for QBCDs. 

Thus, the application of inadequate abundance calibrators to two different samples of objects leads to the erroneous conclusion the the two samples have different metallicities. We believe that it is exactly what is happening in the case of the analysis by S{\'a}nchez Almeida et al. (2008, 2009). 

Why is it so? QBCDs, having older ionizing stellar populations than BCDs, are ionized by a softer radiation field. In addition, their HII regions are presumably less compact and more filamentary because they have suffered the mechanical action of stellar winds for a longer time than BCDs. As a consequence, their ionization parameter is smaller. Both reasons contribute to lowering the mean excitation of QBCD HII regions with respect to BCD ones.  All the calibrators used here (except  O$_{23}$, in panel a) are strongly dependent on the ionization state of the gas. Therefore, they cannot serve to compare the metallicities of two samples which systematically differ in their ionization state! We note that, unfortunately, all the calibrators are biased in the same direction, therefore using several calibrators will not solve the problem. 

The O$_{23}$ calibrator does not depend strongly on the ionization state, but it does depend on the hardness of the heating radiation. Therefore,   if is not fully secure either (and, as is known, it is double valued which makes its use complicated). 

\section{Conclusion}

Strong line methods to derive nebular abundances are very easy to apply but they are prone to systematic errors. In principle, they should be used only for objects  whose HII regions have the  same structural proprties as those of the calibrating samples. This recommendation is not easy to follow, but at least one should be aware that using the same calibration for different samples may produces important biases.

In particular,  claims on differences in oxygen abundance 
\begin{itemize}
\item between samples of galaxies with different chemical evolution histories
\item between samples of galaxies with different star formation histories
\item  between samples of galaxies at different redshifts (observational selection effects may play a role)
\end{itemize}
should be taken with, at a minimum, a grain of salt.

\begin{acknowledgements}

I am very grateful to Yuri Izotov for allowing me to use his measurements of line intensities in the SDSS DR7 galaxies.

\end{acknowledgements}


\begin{thebibliography}{}

\bibitem[]{}Adelman-McCarthy J.~K., et al., 2008, ApJS, 175, 297 
\bibitem[]{}Alloin D., Collin-Souffrin S., Joly M., Vigroux L., 1979, A\&A 78, 200
\bibitem[]{}Dopita M.~A., et al., 2006, ApJ, 647, 244 

\bibitem[]{}Izotov Y.~I., Stasi{\'n}ska G., Meynet G. et al., 2006, A\&A, 448, 955 
\bibitem[]{}Kewley L.~J., Ellison S.~L., 2008, ApJ, 681, 1183

\bibitem[]{}McGaugh S.~S., 1991, ApJ, 380, 140 
\bibitem[]{}Nagao T., Maiolino R., Marconi A., 2006, A\&A, 459, 85 

\bibitem[]{}P{\'e}rez-Montero E., D{\'{\i}}az A.~I., 2005, MNRAS, 361, 1063 
\bibitem[]{}Pagel B.~E.~J., Edmunds M.~G., Blackwell D.~E. et al., 1979, MNRAS, 189, 95 
\bibitem[]{}S{\'a}nchez Almeida J., Aguerri J.~A.~L., Mu{\~n}oz-Tu{\~n}{\'o}n C., Vazdekis A., 2009, ApJ, 698, 1497 
\bibitem[]{}S{\'a}nchez Almeida J., 
Mu{\~n}oz-Tu{\~n}{\'o}n C., Amor{\'{\i}}n R., Aguerri J.~A., 
S{\'a}nchez-Janssen R., Tenorio-Tagle G., 2008, ApJ, 685, 194 



\bibitem[]{}Stasi{\'n}ska G., 2004, cmpe.conf, 115
\bibitem[]{}Stasi{\'n}ska G., 2006, A\&A, 454, L127 
\bibitem[]{}Stasi{\'n}ska G., 2008, IAUS, 255, 375 

\bibitem[]{}Storchi-Bergmann T., Calzetti D., Kinney A.~L., 1994, ApJ, 429, 572 
\bibitem[]{}Vale Asari N., Stasi{\'n}ska G., Cid Fernandes R., Gomes J.~M., Schlickmann M., Mateus A., Schoenell W., 2009, MNRAS, 396, L71 
\bibitem[]{}Vilchez J.~M., Esteban C., 1996, MNRAS, 280, 720 
\end{thebibliography}
\end{document}